\documentclass[a4paper,11pt]{article}
\usepackage[dvipdfmx]{graphicx}
\usepackage{amsfonts}
\usepackage{physics,amsmath}
\usepackage{amssymb}
\usepackage{mathtools}
\usepackage{colortbl}
\usepackage{cite}
\usepackage{here}
\usepackage{hyperref}
\usepackage{mathrsfs}
\usepackage{algpseudocode}
\usepackage{xcolor}
\usepackage{ulem}
\usepackage{comment}
\numberwithin{equation}{section}
\usepackage[left=2cm,right=2cm,top=3.5cm,bottom=3.5cm]{geometry}

\allowdisplaybreaks[1]
\hypersetup{
	colorlinks=true,
	linkcolor=ceruleanblue,
	filecolor=ceruleanblue,      
	urlcolor=ceruleanblue,
	citecolor=ceruleanblue,
}
\definecolor{ceruleanblue}{rgb}{0.0, 0.2, 0.6}

\newcommand{\de}{{\rm d}}

\let\originalleft\left
\let\originalright\right
\renewcommand{\left}{\mathopen{}\mathclose\bgroup\originalleft}
\renewcommand{\right}{\aftergroup\egroup\originalright}

\newcommand{\brb}[1]{\left[ #1 \right]}  
  
\newcommand{\be}{\begin{equation}}  
\newcommand{\ee}{\end{equation}}
\newcommand{\bem}{\begin{bmatrix}}
\newcommand{\eem}{\end{bmatrix}}

\date{\today}

\begin{document}

\begin{flushright} {\footnotesize YITP-25-98, IPMU25-0036}  \end{flushright}

\begin{center}
\LARGE{\bf Effective Field Theory of Perturbations on Arbitrary Black Hole Backgrounds with Spacelike Scalar Profile}
\\[1cm] 

\large{Shinji Mukohyama$^{\,\rm a,b,c}$, Emeric Seraille$^{\,\rm d,e}$, Kazufumi Takahashi$^{\,\rm f,a}$, \\ and Vicharit Yingcharoenrat$^{\,\rm g,c}$}
\\[0.3cm]

\small{
\textit{$^{\rm a}$
Center for Gravitational Physics and Quantum Information, Yukawa Institute for Theoretical Physics, 
\\ Kyoto University, 606-8502, Kyoto, Japan}}
\vspace{.2cm}

\small{
\textit{$^{\rm b}$
Research Center for the Early Universe (RESCEU), Graduate School of Science, The University of Tokyo, Hongo 7-3-1, Bunkyo-ku, Tokyo
113-0033, Japan}}
\vspace{.2cm}

\small{
\textit{$^{\rm c}$
Kavli Institute for the Physics and Mathematics of the Universe (WPI), The University of Tokyo Institutes for Advanced Study (UTIAS), The University of Tokyo, Kashiwa, Chiba 277-8583, Japan}}
\vspace{.2cm}

\small{
\textit{$^{\rm d}$
Laboratoire de Physique de l’Ecole Normale Sup{\'e}rieure, ENS, CNRS, Universit{\'e}  PSL, Sorbonne Universit{\'e}, Universit{\'e} Paris Cit{\'e}, F-75005 Paris, France}}
\vspace{.2cm}

\small{
\textit{$^{\rm e}$
$\mathcal{G}\mathbb{R}\varepsilon{\mathbb{C}}\mathcal{O}$, 
	Institut d'Astrophysique de Paris, UMR 7095, CNRS, Sorbonne Universit{\'e},
	98\textsuperscript{bis} boulevard Arago, 75014 Paris, France}}
\vspace{.2cm}

\small{
\textit{$^{\rm f}$
Department of Physics, College of Humanities and Sciences, Nihon University, Tokyo 156-8550, Japan}}
\vspace{.2cm}

\small{
\textit{$^{\rm g}$
High Energy Physics Research Unit, Department of Physics, Faculty of Science, Chulalongkorn University, Pathumwan, Bangkok 10330, Thailand}}
\vspace{.2cm}
\end{center}

\vspace{0.2cm} 

\begin{abstract}\normalsize
We develop the effective field theory (EFT) of perturbations in the context of scalar-tensor theories with a spacelike scalar profile on arbitrary black hole backgrounds.
Our construction of the EFT is based on the fact that in the unitary gauge, where the scalar field is chosen as one of the spatial coordinates, the background scalar field spontaneously breaks the diffeomorphism invariance along the direction of its gradient.
The residual symmetry on a timelike hypersurface of constant scalar field is referred to as the $(2+1)$d diffeomorphism invariance.
We then derive a set of consistency relations, imposed on the EFT parameters, by requiring that the EFT action in the unitary gauge be invariant under the $(2+1)$d diffeomorphisms.
For concreteness, we apply the EFT to study the background dynamics of a class of non-static and spherically symmetric solutions, focusing in particular on a black hole solution with a time-varying mass.
We emphasize that our EFT framework is broadly applicable to any black hole background as long as the scalar field remains spacelike throughout the spacetime region of interest.
This formulation provides a model-independent approach for testing scalar-tensor theories as gravity beyond general relativity in the strong-gravity regime.
\end{abstract}

\vspace{0.3cm} 

\vspace{2cm}

\newpage
{
\hypersetup{linkcolor=black}
\tableofcontents
}

\flushbottom

\vspace{1cm}

\section{Introduction}\label{sec:intro}
Gravitational-wave (GW) detections by LIGO-Virgo-KAGRA collaborations~\cite{LIGOScientific:2016aoc,LIGOScientific:2018mvr,LIGOScientific:2019lzm,KAGRA:2023pio} have ushered in a new era for testing theories of gravity in the strong-field regime. 
Up to now, many observations appear to be well consistent with general relativity (GR).
However, as more observational data become available, it is important to quantitatively characterize the extensions or deviations from GR through which theories beyond GR can be tested and potentially ruled out (see, e.g., \cite{Koyama:2015vza,Ferreira:2019xrr,Arai:2022ilw}).
In this way, one gains a better understanding of gravity within the regime accessible to observations. 
Moreover, introducing modifications of gravity offers a potential approach to address several mysteries of our Universe---such as dark energy, inflation, and dark matter---that GR alone cannot fully explain.

In the present paper, we focus on the simplest modification of GR: scalar-tensor theories which introduce an additional scalar degree of freedom in addition to the standard two graviton polarizations.
Typical examples of scalar-tensor theories include the Brans-Dicke theories~\cite{Brans:1961sx} and the k-essence theories~\cite{Armendariz-Picon:2000nqq,Armendariz-Picon:2000ulo}.
It was shown that Horndeski theory~\cite{Horndeski:1974wa,Deffayet:2011gz,Kobayashi:2011nu} is the most general class of scalar-tensor theories with covariant, second-order Euler-Lagrange equations in four dimensions, and there have been extensive studies of cosmology and black holes in this framework (see, e.g., \cite{Clifton:2011jh,Babichev:2016rlq} for reviews and references therein).
Following the discovery of some examples of beyond Horndeski theories~\cite{Zumalacarregui:2013pma,Gleyzes:2014dya}---that is, theories with higher-order Euler-Lagrange equations but free from the Ostrogradsky ghost~\cite{Woodard:2015zca}, it was recognized that the so-called degeneracy conditions~\cite{Motohashi:2014opa,Langlois:2015cwa,Motohashi:2016ftl,Klein:2016aiq} play a crucial role in the systematic construction of such theories.
Theories obtained by imposing the degeneracy conditions are known as degenerate higher-order scalar-tensor (DHOST) theories~\cite{Langlois:2015cwa,Crisostomi:2016czh,BenAchour:2016fzp} (see \cite{Langlois:2018dxi,Kobayashi:2019hrl} for comprehensive reviews).
Moreover, demanding that the degeneracy conditions are satisfied in the unitary gauge led to an extension called U-DHOST theories~\cite{DeFelice:2018ewo,DeFelice:2021hps,DeFelice:2022xvq}.
Additionally, performing a higher-order generalization of invertible disformal transformations~\cite{Takahashi:2021ttd,Takahashi:2023vva} on Horndeski theories and U-DHOST theories yields the generalized disformal Horndeski theories~\cite{Takahashi:2022mew} and the generalized disformal unitary-degenerate theories~\cite{Takahashi:2023jro} (see \cite{Takahashi:2022mew,Naruko:2022vuh,Takahashi:2022ctx,Ikeda:2023ntu,Takahashi:2023vva} for discussions on the consistency of matter coupling in these theories). 

In principle, given a symmetry-breaking pattern and a background configuration, all these covariant theories can be described within a unified framework, known as effective field theory (EFT). 
An EFT of scalar-tensor theories on Minkowski/de Sitter background with a timelike scalar profile, also known as the EFT of ghost condensation, was originally formulated in \cite{Arkani-Hamed:2003pdi,Arkani-Hamed:2003juy} based on the assumption that temporal diffeomorphism invariance is spontaneously broken by the timelike gradient of a scalar field.
In that case, the EFT action in the unitary gauge is left invariant under spatial diffeomorphisms.
A similar approach was applied to cosmology, where the background metric is described by the Friedmann-Lema\^itre-Robertson-Walker metric, resulting in the EFT of inflation/dark energy~\cite{Cheung:2007st,Gubitosi:2012hu} (see \cite{Finelli:2018upr} for a shift-symmetric scalar-tensor EFT).
A further extension of such an EFT with a timelike scalar profile was achieved in \cite{Mukohyama:2022enj,Mukohyama:2022skk}, which applies to an arbitrary background metric (see also \cite{Khoury:2022zor} for the EFT on black hole backgrounds with shift symmetry).
In that case, in order for the EFT action to be invariant under the spatial diffeomorphisms, one is required to impose a set of consistency relations among the EFT parameters, derived by applying the chain rule associated with the spatial derivatives to the Taylor coefficients (see \cite{Mukohyama:2022enj,Mukohyama:2022skk,Mukohyama:2023xyf} for detailed discussions).
The studies of linear black hole perturbations based on the EFT, both in the odd- and even-parity sectors, can be found in \cite{Mukohyama:2022skk,Mukohyama:2025owu}.
It is worth noting that the analysis in \cite{Mukohyama:2025owu} was carried out on an approximately stealth Schwarzschild solution~\cite{DeFelice:2022qaz}, incorporating the so-called scordatura term~\cite{Motohashi:2019ymr}.
Phenomenological aspects of the EFT have also been extensively studied, including the quasinormal mode spectrum~\cite{Mukohyama:2023xyf,Konoplya:2023ppx}, graybody factors~\cite{Konoplya:2023ppx,Oshita:2024fzf}, tidal Love numbers~\cite{Barura:2024uog}, and tidal dissipation numbers~\cite{Kobayashi:2025swn}.
See also \cite{Mukohyama:2024pqe} for a discussion on conformal and disformal transformations of the EFT parameters on arbitrary backgrounds.

Apart from the EFT of perturbations with a timelike scalar profile, in \cite{Franciolini:2018uyq}, the EFT of black hole perturbations with a spacelike scalar profile was first formulated on a static and spherically symmetric background.\footnote{See \cite{Kuntz:2020yow} for a detailed analysis of extreme mass ratio inspirals using the EFT approach.}
Later, it was extended to accommodate a slowly rotating background in \cite{Hui:2021cpm}. 
It should be pointed out that under certain assumptions---e.g., if the background is asymptotically flat---a non-trivial scalar profile is forbidden by the no-hair theorem \cite{Bekenstein:1971hc,Bekenstein:1995un,Hui:2012qt,Capuano:2023yyh}.
(See also \cite{Takahashi:2020hso,Kobayashi:2025evr} for recent studies on how requiring GR solutions restricts scalar-tensor theories.)
When some of the no-hair assumptions are violated, it is then possible to obtain non-trivial hairy solutions (see, e.g., \cite{Dennhardt:1996cz,Mukohyama:2005rw,Babichev:2013cya,Sotiriou:2014pfa,Antoniou:2017hxj}).
In fact, it was shown that, in shift-symmetric Horndeski and DHOST theories, the scalar-Gauss-Bonnet coupling is necessary to generate a non-trivial scalar hair, under the assumptions that the scalar profile is time-independent and the background metric is asymptotically flat~\cite{Sotiriou:2015pka,Blazquez-Salcedo:2017txk,Creminelli:2020lxn,Minamitsuji:2022mlv,Babichev:2024txe}.
In this paper, we generalize the EFT formulation developed in \cite{Franciolini:2018uyq,Hui:2021cpm} to an arbitrary background metric, following the EFT approach with a timelike scalar profile in \cite{Mukohyama:2022enj,Mukohyama:2022skk}.
Our EFT in principle applies to general stationary and axisymmetric black holes, without being restricted to the slowly rotating case.
In addition to black hole physics, our EFT can be used to study dynamics of perturbations in other contexts such as braneworld scenarios where the gradient of the bulk scalar field is spacelike~\cite{Goldberger:1999uk,Bazeia:2008zx,Dzhunushaliev:2009va}.

The remainder of this paper is organized as follows.
In Section~\ref{sec:formulation_EFT}, we formulate the EFT of perturbations on arbitrary black hole backgrounds with a spacelike scalar profile. 
We also derive a set of consistency relations that are imposed by the invariance of the EFT action under the $(2+1)$d diffeomorphisms on a timelike hypersurface of constant scalar field.
In Section~\ref{sec:bg_dynamics}, we analyze the background dynamics on a spherically symmetric metric that is not necessarily static.
In Section~\ref{sec:example_slow_BH}, we apply the framework to a spherically symmetric black hole solution with a linearly time-varying mass.
Finally, we draw our conclusions and discuss future directions in Section~\ref{sec:conc}.

\section{General formulation of the EFT}\label{sec:formulation_EFT}
\subsection{EFT building blocks}\label{sec:uni_gen_EFT}
In this section, we formulate the EFT of scalar-tensor theories, where the scalar field~$\Phi$ is assumed to have a spacelike profile.
The construction of the EFT is based on the fact that the background scalar field~$\bar{\Phi}$ spontaneously breaks the diffeomorphism invariance along the direction of its gradient.
Due to the spacelike nature of $\Phi$, one can introduce a spacelike coordinate~$r$ such that $\bar{\Phi}=\bar{\Phi}(r)$ and $\delta\Phi=0$, which corresponds to working in the unitary gauge.
In what follows, we refer to $r$ as the radial coordinate, such that the orthogonal directions correspond to temporal and angular coordinates.
In the unitary gauge, the EFT action must be invariant under the $(2+1)$d diffeomorphisms on a timelike hypersurface of constant $\Phi$, whose unit normal vector is defined by
\begin{equation}\label{eq:normal_vector}
    n_{\mu} \equiv \frac{\nabla_{\mu}{\Phi}}{\sqrt{\nabla^{\alpha}{\Phi}\nabla_{\alpha}{\Phi}}} \rightarrow \frac{\delta_{\mu}^{r}}{\sqrt{g^{r r}}} \;.
\end{equation}
Here, the arrow denotes the expression in the unitary gauge, and $\nabla_\mu$ is the 4d covariant derivative.
Note that $n_\mu$ is a spacelike vector such that $n^{\mu}n_{\mu}=1$, and therefore the projection tensor is defined as $h_{\mu\nu}\equiv g_{\mu\nu}-n_\mu n_\nu$.
We introduce the Arnowitt-Deser-Misner (ADM) decomposition of the metric with respect to the radial coordinate~$r$ as follows:
\begin{align}
    g_{\mu\nu}{\rm d}x^\mu{\rm d}x^\nu=N^{2} \de r^2 +h_{ab}\bigl(\de x^{a} +N^{a}\de r\bigr)\bigl(\de x^{b} + N^{b}\de r\bigr)\;,
\end{align}
where the indices~$a$ and $b$ refer to the temporal and angular coordinates~$\{t,\theta, \phi\}$, $N$ is the lapse function, $N^{a}$ is the shift vector, and $h_{ab}$ is the induced metric on a constant-$\Phi$ hypersurface. Written in terms of the ADM quantities, the components of the metric and its inverse are given by
\begin{align}
    g_{r r} = N^2 + h_{a b}N^{a} N^{b} \;, \qquad g_{a r} =h_{a b}N^{b} \equiv N_{a} \;, \qquad g_{a b}= h_{a b} \;,
\end{align}
and
\begin{align}
    g^{r r} = \frac{1}{N^2} \;, \qquad g^{a r} = -\frac{N^{a}}{N^2} \;, \qquad g^{a b}= h^{a b}+ \frac{N^{a}N^{b}}{N^2} \;,
\end{align}
respectively.
Using the projection tensor, we write the extrinsic curvature associated with the foliation as
\begin{equation}
K_{\mu \nu} \equiv h^{\sigma}_{\mu}\nabla_{\sigma}n_{\nu} \;,
\end{equation}
and its trace as $K\equiv K^\mu_\mu$.
Written in terms of the ADM variables,
\begin{equation}
K_{a b}=\frac{1}{2N}\bigl(\partial_{r}{h}_{ab}-{\rm D}_{a}N_{b}-{\rm D}_{b}N_{a}\bigr)  \;, \qquad K= h^{a b}K_{a b} \;,
\end{equation}
where ${\rm D}_{a}$ is the covariant derivative associated with the induced metric.
Additionally, we can construct any $(2+1)$d geometrical quantities such as the $(2+1)$d Ricci scalar~${}^{(3)}\!R$, out of the induced metric and its derivatives in the standard manner. 
It is useful to note that the Gauss-Codazzi equation in this case reads 
\begin{align}\label{eq:gauss_codazzi}
    h^\alpha_\mu h^\beta_\nu h^\gamma_\rho h^\delta_\sigma R_{\alpha\beta\gamma\delta}
    = {}^{(3)}\!R_{\mu\nu\rho\sigma}+K_{\mu\sigma} K_{\nu\rho}-K_{\mu\rho} K_{\nu\sigma} \;,
\end{align}
where $R_{\alpha\beta\gamma\delta}$ and ${}^{(3)}\!R_{\mu\nu\rho\sigma}$ are the 4d and the $(2+1)$d curvature tensors, respectively.

As explained in \cite{Cheung:2007st,Franciolini:2018uyq,Mukohyama:2022enj}, in addition to the 4d covariant quantities such as the 4d Ricci scalar, the unitary-gauge action can contain scalar functions that are made out of the $(2+1)$d diffeomorphism covariant quantities, for instance, $N$, $K_{\mu\nu}$, and ${}^{(3)}\!R_{\mu\nu}$.
Note that, thanks to the Gauss-Codazzi equation~(\ref{eq:gauss_codazzi}), one can always replace the 4d curvature tensor and their contractions with the $(2+1)$d curvature tensor.
Also, since the Weyl tensor vanishes identically in 3d, we can use the $(2+1)$d Ricci tensor~${}^{(3)}\!R_{\mu\nu}$ as an independent building block.
In addition, an explicit function of $r$ can also be included in the EFT action.
Given all possibilities explained above, the most general unitary-gauge EFT action is given by
\begin{align}\label{eq:EFT_unitary}
    S = \int {\rm d}^4 x \sqrt{-g}\,\mathcal{L}(g^{rr}, K^\mu_{\nu}, {}^{(3)}\!R^\mu_\nu, \nabla_\mu, r) \;,
\end{align}
where $\mathcal{L}$ is an arbitrary scalar function which is invariant under the $(2+1)$d diffeomorphisms.
We emphasize that the EFT action above can be applied to arbitrary background geometries without assuming any symmetries. 
In order to apply this EFT action to a specific system, we will proceed by expanding it around a non-trivial background.

To do so, let us define perturbations of the EFT building blocks as 
\begin{align}\label{eq:building_block}
    \delta g^{rr} \equiv g^{rr} - \bar{g}^{rr}(x) \;, \qquad \delta K^\mu_\nu \equiv K^\mu_\nu - \bar{K}^\mu_\nu(x) \;, \qquad \delta{}^{(3)}\!R^\mu_\nu \equiv {}^{(3)}\!R^\mu_\nu - {}^{(3)}\!\bar{R}^\mu_\nu(x) \;, 
\end{align}
where a bar denotes the background quantity.
Notice that the background values, introduced above, can be functions of space and time. 
We will see, later on, that the dependence of those background values on space and time coordinates explicitly breaks the $(2+1)$d diffeomorphism invariance at the level of individual terms, and therefore one needs to impose the consistency relations among the EFT coefficients so that the EFT action as a whole preserves the invariance.

More explicitly, performing the Taylor expansion of the action~(\ref{eq:EFT_unitary}) around a non-trivial background metric gives 
\begin{align}
    S &= \int {\rm d}^4x\sqrt{-g}\bigg[\bar{\mathcal{L}}(x) + \bar{\mathcal{L}}_{g^{rr}}(x) \delta g^{rr} + \bar{\mathcal{L}}_{K}(x) \delta K + \bar{\mathcal{L}}_{\sigma^\mu_\nu}(x) \delta \sigma^\mu_\nu + \bar{\mathcal{L}}_{{}^{(3)}\!R}(x) \delta {}^{(3)}\!R + \bar{\mathcal{L}}_{\tilde{r}^\mu_\nu}(x) \delta \tilde{r}^\mu_\nu + \nonumber \\ 
    & \hspace{4mm} + \frac{1}{2}\bar{\mathcal{L}}_{g^{rr}g^{rr}}(x)(\delta g^{rr})^2 + \bar{\mathcal{L}}_{g^{rr}K}(x) \delta g^{rr} \delta K + \frac{1}{2}\bar{\mathcal{L}}_{KK}(x) \delta K^2 + \frac{1}{2}\bar{\mathcal{L}}_{\sigma^2}(x) \delta \sigma^\mu_\nu \delta \sigma^\nu_\mu + \cdots \bigg] \;, \label{eq:EFT_taylor}
\end{align}
where we have defined $\bar{\mathcal{L}}_Q \equiv (\partial \mathcal{L}/\partial Q)_{\rm bkg}$, $\sigma^\mu_\nu \equiv K^\mu_\nu - K h^\mu_\nu/3$, and $\tilde{r}^\mu_\nu \equiv {}^{(3)}\!R^\mu_\nu - {}^{(3)}\!R\, h^\mu_\nu/3$.
In Eq.~(\ref{eq:EFT_taylor}), we have included all terms linear in perturbations (i.e., tadpole terms), whereas keeping only a few terms at second order in perturbations.\footnote{The ellipsis refers to those terms we did not write down as well as the higher-order terms in perturbations and/or derivatives.}
Notice that, for a background metric where $\bar{K}^\mu_\nu$ and ${}^{(3)}\!\bar{R}^\mu_\nu$ are proportional to $\bar{h}^\mu_\nu$, the traceless tensors~$\bar{\sigma}^\mu_\nu$ and $\bar{\tilde{r}}^{\mu}_\nu$ vanish.
As we see in Appendix~\ref{AppA:background_Sch}, for the Schwarzschild background, both $\bar{\sigma}^\mu_\nu$ and $\bar{\tilde{r}}^{\mu}_\nu$ have non-vanishing components.
Additionally, it should be emphasized that in Eq.~(\ref{eq:EFT_taylor}) all Taylor coefficients, arising from the expansion of the EFT Lagrangian around a non-trivial background, can in general depend on both space and time.

Let us now comment on the EFT action~(\ref{eq:EFT_taylor}).
Although we formulate here the EFT of perturbations with a spacelike scalar profile, the story is indeed similar to the case with a timelike scalar profile~\cite{Mukohyama:2022enj,Mukohyama:2022skk}.
From Eq.~(\ref{eq:EFT_taylor}), it is evident that each individual term of the Taylor expansion breaks the $(2+1)$d diffeomorphism invariance in general.
This is due to the fact that the perturbations of the EFT building blocks defined in Eq.~(\ref{eq:building_block}) do not transform covariantly under the $(2+1)$d diffeomorphisms, as the background values can depend explicitly on the time and angular coordinates. 
However, as discussed earlier and in the next subsection, the requirement that the full unitary-gauge EFT action is invariant under the $(2+1)$d diffeomorphisms implies that there must be a set of consistency relations.
For similar discussions for the EFT with a timelike scalar profile, see \cite{Mukohyama:2022enj,Mukohyama:2022skk,Mukohyama:2024pqe}.\footnote{Actually, the use of consistency relations was also presented in the case of EFT of vector-tensor gravity on a cosmological background~\cite{Aoki:2021wew} and on arbitrary backgrounds~\cite{Aoki:2023bmz}.}

\subsection{Consistency relations for 2+1 diffeomorphism}\label{sec:Consistency}
Following the same procedure as in \cite{Mukohyama:2022enj,Mukohyama:2022skk}, the consistency relations for the $(2+1)$d diffeomorphism invariance can be obtained by using the chain rule associated with the $(2+1)$d coordinates~$x^a$.
At the leading order in perturbations, applying the chain rule to $\bar{\mathcal{L}}$ defined in Eq.~(\ref{eq:EFT_taylor}) yields
\begin{align}
    \frac{\partial}{\partial x^a} \bar{\mathcal{L}}(r, x^b)&= \frac{{\rm d}}{{\rm d}x^a} \mathcal{L}(g^{rr}, K^\mu_\nu, {}^{(3)}\!R^\mu_\nu, \nabla_\mu, r) \bigg|_{\rm bkg} \nonumber \\
    &= \bar{\mathcal{L}}_{g^{rr}} \partial_a \bar{g}^{rr} + \bar{\mathcal{L}}_{K} \partial_a \bar{K} + \bar{\mathcal{L}}_{\sigma^\mu_\nu} \partial_a \bar{\sigma}^\mu_\nu + \bar{\mathcal{L}}_{{}^{(3)}\!R} \partial_a {}^{(3)}\!\bar{R} + \bar{\mathcal{L}}_{\tilde{r}^\mu_\nu} \partial_a \bar{\tilde{r}}^\mu_\nu + \cdots \;, \label{eq:consistency_tadpole}
\end{align}
where the ellipsis refers to terms of higher order in derivatives. 
From the above relation, we see that the zeroth-order Taylor coefficient~$\bar{\mathcal{L}}$ is related to those at the first order in perturbations; they are not independent of each other.
Clearly, for a static and spherically symmetric background where all the background quantities depend only on $r$, the above consistency relation is satisfied trivially. 
It should be noted that the chain rule associated with the radial direction does not provide non-trivial relations among the Taylor coefficients because each coefficient is allowed to have an explicit dependence on $r$. 
For example, applying the chain rule associated with the $r$ coordinate to $\bar{\mathcal{L}}$ gives 
\begin{align}
    \frac{\partial}{\partial r} \bar{\mathcal{L}}(r, x^b)&= \frac{{\rm d}}{{\rm d}r} \mathcal{L}(g^{rr}, K^\mu_\nu, {}^{(3)}\!R^\mu_\nu, \nabla_\mu, r) \bigg|_{\rm bkg} \nonumber \\
    &= \partial_r\bar{\mathcal{L}}  + \bar{\mathcal{L}}_{g^{rr}} \partial_r \bar{g}^{rr} + \bar{\mathcal{L}}_{K} \partial_r \bar{K} + \bar{\mathcal{L}}_{\sigma^\mu_\nu} \partial_r \bar{\sigma}^\mu_\nu + \bar{\mathcal{L}}_{{}^{(3)}\!R} \partial_r {}^{(3)}\!\bar{R} + \bar{\mathcal{L}}_{\tilde{r}^\mu_\nu} \partial_r \bar{\tilde{r}}^\mu_\nu + \cdots \;. \label{eq:con_x_a_1}
\end{align}
Since $\partial_r\bar{\mathcal{L}}$ does not appear in the EFT action, one can always choose $\partial_r\bar{\mathcal{L}}$ such that the consistency relation above is automatically satisfied for given values of the EFT parameters. 
For this reason, the relations associated with the $r$-derivative do not give non-trivial relations among the EFT coefficients. 
However, in the shift-symmetric EFT where an explicit $r$-dependence is forbidden, the chain rule associated with the 
$r$-derivative leads to non-trivial relations among the EFT parameters. 
We will discuss this point in detail in Section~\ref{sec:shift-sym_EFT}.

In addition, apart from Eq.~(\ref{eq:consistency_tadpole}), there are infinitely many consistency relations. 
For instance, the chain rule applied to $\bar{\mathcal{L}}_{g^{rr}}$ and $\bar{\mathcal{L}}_{K}$ gives
\begin{align}
    \frac{\partial }{\partial x^a}\bar{\mathcal{L}}_{g^{rr}}(r, x^b) &= \bar{\mathcal{L}}_{g^{rr} g^{rr}} \partial_a \bar{g}^{rr} + \bar{\mathcal{L}}_{g^{rr} K} \partial_a \bar{K} + \cdots \;, \label{eq:consis_2} \\ 
    \frac{\partial }{\partial x^a}\bar{\mathcal{L}}_{K}(r, x^b) &= \bar{\mathcal{L}}_{g^{rr} K} \partial_a \bar{g}^{rr} + \bar{\mathcal{L}}_{K K} \partial_a \bar{K} + \cdots \;, \label{eq:consis_3} 
\end{align}
where again the ellipsis refers to terms of higher order in derivatives, and we have explicitly written only the terms that are present in Eq.~(\ref{eq:EFT_taylor}).\footnote{It is in fact straightforward to take into account other second-order terms, as already discussed in detail in \cite{Mukohyama:2022enj,Mukohyama:2022skk} for the EFT of scalar-tensor gravity with a timelike scalar profile.}
One can straightforwardly apply a similar chain rule to other coefficients in the EFT action such as $\bar{\mathcal{L}}_{g^{rr} g^{rr}}$ and $\bar{\mathcal{L}}_{g^{rr} K}$.
At this stage, we impose the consistency relations~(\ref{eq:consistency_tadpole}), (\ref{eq:consis_2}), and (\ref{eq:consis_3}) along with those for other coefficients on the EFT action~(\ref{eq:EFT_taylor}) to ensure that the $(2+1)$d diffeomorphism invariance of the EFT is preserved up to the order of our interest.

\subsection{Consistency relations for shift symmetry (optional)}\label{sec:shift-sym_EFT}
As discussed earlier, the consistency relations we have derived in the previous subsection are sufficient to ensure that the EFT action~\eqref{eq:EFT_taylor} is invariant under the $(2+1)$d diffeomorphism.
In addition, one may optionally impose invariance of the EFT under the shift of the scalar field, $\Phi \to \Phi+const$.\footnote{See \cite{Finelli:2018upr} for the shift-symmetric EFT in cosmology, and \cite{Khoury:2022zor} for the shift-symmetric EFT of black hole perturbations with a timelike scalar profile.}
In the language of covariant theories, the shift symmetry forbids the explicit dependence on $\Phi$.
This implies that, in the unitary gauge where $\bar{\Phi}$ is a fixed function of $r$, the EFT action~\eqref{eq:EFT_unitary} cannot depend on the coordinate~$r$ explicitly, i.e.,
\begin{align}\label{eq:EFT_unitary_shift}
    S = \int {\rm d}^4 x \sqrt{-g}\,\mathcal{L}(g^{rr}, K^\mu_{\nu}, {}^{(3)}\!R^\mu_\nu, \nabla_\mu) \;.
\end{align}

When expanding the action~(\ref{eq:EFT_unitary_shift}) around a non-trivial background, we obtain an expression similar to Eq.~(\ref{eq:EFT_taylor}). 
However, the fact that the background quantities such as $\bar{g}^{rr}$ and $\bar{K}$ generally depend on all spacetime coordinates implies that the Taylor coefficients can inherit the $r$-dependence.
This means that each individual term in the Taylor expansion explicitly breaks the shift symmetry. 
Similarly to the previous subsection, since the action as a whole is invariant under the shift symmetry, there must be a set of consistency relations among the Taylor coefficients associated with the symmetry. 
Such a set of consistency relations can be derived by applying the chain rule associated with $r$ derivatives to the Taylor coefficients. 
Written explicitly, for the Taylor coefficient~$\bar{\mathcal{L}}(r, x^a)$, we have
\begin{align}
    \frac{\partial}{\partial r}\bar{\mathcal{L}}(r, x^b) &= \frac{{\rm d}}{{\rm d}r} \mathcal{L}(g^{rr}, K^\mu_\nu, {}^{(3)}\!R^\mu_\nu, \nabla_\mu) \bigg|_{\rm bkg} \nonumber \\
    &= \bar{\mathcal{L}}_{g^{rr}} \partial_r \bar{g}^{rr} + \bar{\mathcal{L}}_{K} \partial_r \bar{K} + \bar{\mathcal{L}}_{\sigma^\mu_\nu} \partial_r \bar{\sigma}^\mu_\nu + \bar{\mathcal{L}}_{{}^{(3)}\!R} \partial_r {}^{(3)}\!\bar{R} + \bar{\mathcal{L}}_{\tilde{r}^\mu_\nu} \partial_r \bar{\tilde{r}}^\mu_\nu + \cdots \;, \label{eq:shift_con1}
\end{align}
where the ellipsis denotes terms of higher order in derivatives. 
Notice that, compared to Eq.~\eqref{eq:con_x_a_1}, the function~$\partial_r \bar{\mathcal{L}}$ is absent in the above consistency relation.
For other Taylor coefficients in Eq.~(\ref{eq:EFT_taylor}), we obtain 
\begin{align}
    \frac{\partial }{\partial r}\bar{\mathcal{L}}_{g^{rr}}(r, x^b) &= \bar{\mathcal{L}}_{g^{rr} g^{rr}} \partial_r \bar{g}^{rr} + \bar{\mathcal{L}}_{g^{rr} K} \partial_r \bar{K} + \cdots \;, \label{eq:shift_con2} \\
    \frac{\partial }{\partial r} \bar{\mathcal{L}}_{K}(r, x^b) &= \bar{\mathcal{L}}_{g^{rr} K} \partial_r \bar{g}^{rr} + \bar{\mathcal{L}}_{K K} \partial_r \bar{K} + \cdots \;. \label{eq:shift_con3}
\end{align}
Imposing the relations~(\ref{eq:shift_con1})--(\ref{eq:shift_con3}) together with those for higher-order terms guarantees that the EFT action is consistent with the shift symmetry. 

\subsection{EFT action}\label{sec:EFT_consis_action}
In the previous subsections, we have discussed the consistency relations that are imposed to ensure the $(2+1)$d diffeomorphism invariance of the EFT as well as those associated with shift symmetry.
Here, we write down the EFT action in the unitary gauge as 
\begin{align}
    S &= \int {\rm d}^4x \sqrt{-g} \bigg[\frac{M^2(x)}{2}R - \Lambda(x) - c(x) g^{rr} - \beta(x) K - \alpha^\mu_\nu(x) \sigma^\nu_\mu - \gamma^\mu_\nu(x) \tilde{r}^\nu_\mu \nonumber \\ 
   & \hspace{4mm} + \frac{1}{2}m_2^4(x) (\delta g^{rr})^2 + \frac{1}{2}M_1^3(x) \delta g^{rr} \delta K + \frac{1}{2}M_2^2(x) \delta K^2 + \frac{1}{2}M_3^2(x) \delta K^\mu_\nu \delta K^\nu_\mu + \cdots \bigg] \;, \label{eq:uni_EFT}
\end{align}
where $R$ is the 4d Ricci scalar with the boundary term subtracted, defined by\footnote{In a general case where $n_\mu n^\mu = \epsilon$, we define
\begin{align*}
    R \equiv {}^{(3)}\!R + \epsilon (K^2 - K^\mu_\nu K^\nu_\mu) = \tilde{R} - 2 \epsilon \nabla_\mu( n^\nu \nabla_\nu n^\mu - K n^\mu) \;.
\end{align*}
The case with a timelike scalar profile can be recovered when $\epsilon = -1$.
}
\begin{align}
    R \equiv {}^{(3)}\!R + K^2 - K^\mu_\nu K^\nu_\mu = \tilde{R} - 2 \nabla_\mu( n^\nu \nabla_\nu n^\mu - K n^\mu) \;, \label{def_R}
\end{align}
with $\tilde{R}$ being the standard 4d Ricci scalar.
From Eq.~(\ref{eq:uni_EFT}), terms in the first line are tadpole terms which describe the background dynamics, while terms in the second line affect the dynamics of perturbations. 
In Eq.~(\ref{eq:uni_EFT}), the terms we kept at second order are those corresponding to terms that appear in Eq.~(\ref{eq:EFT_taylor}).
Note that the EFT parameters in Eq.~(\ref{eq:uni_EFT}) are functions of both space and time in general.
It should also be noted that the EFT action has an ambiguity up to total derivatives.
Indeed, for an arbitrary function~$\tilde{G}(r)$, we have
    \begin{align}
    \tilde{G}(r)K=-\tilde{G}'(r)\sqrt{g^{rr}}+\nabla_\mu\left[\tilde{G}(r)n^\mu\right]\;, \label{beta_redef}
    \end{align}
with $\tilde{G}'(r)\equiv {\rm d}\tilde{G}/{\rm d}r$.
This relation implies that an arbitrary function of $r$ extracted from $\beta$ can be absorbed into $\Lambda$, $c$, $m_2^4$, and EFT coefficients associated with cubic- or higher-order terms in $\delta g^{rr}$ which are not explicitly shown in Eq.~\eqref{eq:uni_EFT}.

The EFT parameters, introduced in Eq.~(\ref{eq:uni_EFT}), are related to the Taylor coefficients in Eq.~(\ref{eq:EFT_taylor}) as 
\begin{equation}\label{eq:match_con}
\begin{aligned}
    &M^2 = 2 \bar{\mathcal{L}}_{{}^{(3)}\!R} \;, \quad
    \Lambda = -\bar{\mathcal{L}} + \bar{\mathcal{L}}_{{}^{(3)}\!R} \bigg({}^{(3)}\!\bar{R} - \frac{2}{3} \bar{K}^2 + \bar{\sigma}^\mu_\nu \bar{\sigma}^\nu_\mu\bigg) + \bar{\mathcal{L}}_{g^{rr}} \bar{g}^{rr} +  \bar{\mathcal{L}}_{K} \bar{K} + \bar{\mathcal{L}}_{\sigma^\mu_\nu} \bar{\sigma}^\mu_\nu + \bar{\mathcal{L}}_{\tilde{r}^\mu_\nu} \bar{\tilde{r}}^\mu_\nu \;, \\
    &c = - \bar{\mathcal{L}}_{g^{rr}} \;, \quad \alpha^\mu_\nu = -\bar{\mathcal{L}}_{\sigma^\nu_\mu} - 2 \bar{\mathcal{L}}_{{}^{(3)}\!R} \bar{\sigma}^\mu_\nu \;, \quad \beta = -\bar{\mathcal{L}}_{K} + \frac{4}{3}\bar{\mathcal{L}}_{{}^{(3)}\!R} \bar{K} \;, \\ 
    &m_2^4 = \bar{\mathcal{L}}_{g^{rr}g^{rr}} \;, \quad M_1^3 = 2 \bar{\mathcal{L}}_{g^{rr}K} \;, \quad M_2^2 = \bar{\mathcal{L}}_{KK} - 2 \bar{\mathcal{L}}_{{}^{(3)}\!R} - \frac{1}{3} \bar{\mathcal{L}}_{\sigma^2} \;, \quad M_3^2 = \bar{\mathcal{L}}_{\sigma^2} + 2 \bar{\mathcal{L}}_{{}^{(3)}\!R} \;.
\end{aligned}
\end{equation}
Using the relation above, one can straightforwardly obtain the dictionary between the EFT coefficients and functions involved in covariant theories such as Horndeski theory or higher-order scalar-tensor (HOST) theories.
Then, the consistency relations~(\ref{eq:consistency_tadpole}), (\ref{eq:consis_2}), and (\ref{eq:consis_3}) for the $(2+1)$d diffeomorphism invariance can be rewritten as
\begin{align}
    \partial_a \Lambda + \bar{g}^{rr}\partial_a c - \frac{1}{2}\bar{R}\partial_aM^2 + \bar{\sigma}^\mu_\nu \partial_a \alpha^\nu_\mu + \bar{K}\partial_a \beta + \bar{\tilde{r}}^\mu_\nu \partial_a \gamma^\nu_\mu &= 0 \;, \label{eq:EFT_consis_1} \\
    \partial_a c + m_2^4 \partial_a \bar{g}^{rr} + \frac{1}{2}M_1^2\partial_a \bar{K} &= 0 \;, \label{eq:EFT_consis_2} \\
    \partial_a \beta + \frac{1}{2}M_1^2\partial_a \bar{g}^{rr} + \bigg(M_2^2 + \frac{1}{3} M_3^2\bigg) \partial_a \bar{K} - \frac{2}{3}\bar{K} \partial_a M^2 &= 0 \;. \label{eq:EFT_consis_3}
\end{align}
The optional consistency relations~(\ref{eq:shift_con1})--(\ref{eq:shift_con3}) for the shift symmetry can be rewritten in a similar manner, and they are simply Eqs.~\eqref{eq:EFT_consis_1}--\eqref{eq:EFT_consis_3} with the derivatives with respect to $x^a$ replaced by those with respect to $r$.
In the next section, we will apply these relations to a specific background configuration.

Let us now comment on the advantages of the EFT~(\ref{eq:uni_EFT}) with the consistency conditions~\eqref{eq:EFT_consis_1}--\eqref{eq:EFT_consis_3}.
Recall that the EFT relies only on the existence of a scalar field with a spacelike profile and does not assume any concrete covariant theory.
As we shall see in Section~\ref{sec:example_slow_BH}, once tadpole cancellation conditions are imposed on the EFT to ensure consistency with a chosen background, the EFT can be used directly to study the dynamics of perturbations around that background.\footnote{In fact, this allows us to extract interesting phenomenologies even without specifying explicit functional forms of the background. In the case of the EFT with a timelike scalar profile, it was pointed out in \cite{Mukohyama:2024pqe} that the speeds of GWs in the radial and angular directions are generally different. This phenomenon was obtained without relying on the detailed form of the metric, but only on the general assumption that the background is static and spherically symmetric.}
Using dictionaries, constraints on the EFT parameters from observations or theoretical requirements (e.g., the absence of ghost/gradient instabilities) can also be translated into constraints on a broad class of concrete theories.
(See \cite{Creminelli:2017sry,Creminelli:2018xsv,Creminelli:2019nok,Creminelli:2019kjy} for applications of the EFT framework to (beyond) Horndeski theories.)
Moreover, the EFT framework can in principle encompass theories beyond known classes of concrete theories, allowing for phenomenologies not achievable in existing models.
In this sense, employing the EFT approach is more informative than studying perturbations  around fixed backgrounds in a concrete theory.

It is important to mention that the full 4d diffeomorphism invariance of the EFT can be restored using the so-called Stueckelberg trick.
Such a procedure introduces a Goldstone boson field~$\pi$, which non-linearly realizes the radial diffeomorphism symmetry, via a transformation $r \rightarrow r + \pi(r,x^a)$ (see, e.g., \cite{Franciolini:2018uyq}).\footnote{In the case of a timelike scalar profile, the Stueckelberg procedure for EFT building blocks up to second order in $\pi$ can be found in \cite{Cusin:2017mzw,Mukohyama:2022enj}.} 
This field $\pi$ plays a role of a scalar fluctuation on arbitrary backgrounds. 
Using the transformation~$r \to r + \pi(r,x^a)$, the EFT operators that are invariant only under the $(2+1)$d diffeomorphisms will generate terms involving the field~$\pi(r,x^a)$. 
For example, the operator~$g^{rr}$ gives rise to 
\begin{align}
    g^{rr} \to g^{rr} (1 + 2\pi' + \pi'^2) + 2 g^{ar}\partial_a \pi + 2 g^{ar} \pi' \partial_a \pi + g^{ab} \partial_a \pi \partial_b \pi \;, 
\end{align}
where the arrow here refers to applying the Stueckelberg trick, and a prime stands for the derivative with respect to $r$.
It is in fact straightforward to derive the Stueckelberg trick for other $(2+1)$d quantities such as the lapse function, the shift vector, the extrinsic curvature, and the $(2+1)$d Ricci scalar.\footnote{Note that the quantity~$R$ defined in Eq.~\eqref{def_R} is invariant only under the $(2+1)$d diffeomorphisms, since the subtracted boundary term is not 4d covariant.
Therefore, the Stueckelberg trick should be applied to it.}
Notice that this procedure of introducing the Stueckelberg field~$\pi(r,x^a)$ is particularly useful when studying dynamics of the scalar perturbation in the decoupling limit.

Before closing this section, it is worth pointing out that the EFT we constructed in this paper is valid as long as $\bar{\Phi}'(r)$ is non-vanishing. 
Therefore, the validity of the EFT becomes subtle if the scalar profile approaches a trivial configuration in the asymptotic region, i.e., $\bar{\Phi}'(r) \to 0$.
If this indeed happens, one is then required to construct another EFT around such a trivial background and connect it with the one we have formulated in this paper.\footnote{Roughly speaking, if the strong-coupling scale of our EFT (or the unitarity cutoff) decreases more slowly than the size of perturbations as $r \to \infty$, the EFT would remain valid. This point requires further investigation.}
To perform such a connection, some input from a (partial) UV completion of our EFT may be needed. 
Additionally, it should be emphasized that changing the EFT to another alters the boundary conditions used to solve the equations of motion, and consequently affects observables such as the quasinormal-mode spectrum and tidal Love numbers. 
We will discuss this point further in Section~\ref{sec:conc}.


\section{Spherically symmetric background}\label{sec:bg_dynamics}
The formalism developed in the previous section is applicable to arbitrary backgrounds as far as the gradient of the scalar field is spacelike. In this section, for practical purposes, we derive the background equations of motion (or equivalently, tadpole cancellation conditions) from our EFT action~(\ref{eq:uni_EFT}). 
Note that, throughout the present paper, we only focus on black hole solutions in the absence of matter fields. 
In this section, we assume the background metric to be spherically symmetric but not necessarily static, i.e.,
\begin{align}\label{eq:bg_metric_Le}
	\bar{g}_{\mu\nu}{\rm d}x^\mu{\rm d}x^\nu=- A (t,r)\de t^2 + \frac{\de r^2}{B(t,r)}  + F(t,r) \left(\de {\theta}^2 + {\sin}^2\theta\,\de {\phi}^2\right)\;,
\end{align}
where $A(t,r)$, $B(t,r)$, and $F(t,r)$ are arbitrary functions of the radial and time coordinates.
It should be noted that the coordinate~$r$ is not necessarily the areal radius, and thus $F\ne r^2$ in general.
Indeed, we have already used the coordinate freedom to choose $r$ such that the background scalar field~$\bar{\Phi}$ depends only on $r$, i.e., $\bar{\Phi}=\bar{\Phi}(r)$.
For the time being, we do not assume specific forms of the functions~$A(t,r)$, $B(t,r)$, and $F(t,r)$.
The background quantities, such as $\bar{K}$ and ${}^{(3)}\!\bar{R}$, corresponding to the metric~(\ref{eq:bg_metric_Le}) are given in Appendix~\ref{AppA:background_general}.
We emphasize that the metric~(\ref{eq:bg_metric_Le}) naturally accommodates, for instance, the McVittie metric~\cite{McVittie:1933zz}, which describes a black hole embedded in an expanding universe.
This connection could help illustrate phenomenological applications related to black holes and cosmology. 
We leave further investigation to future work.

For simplicity, we assume $\gamma^{\mu}_{\nu}=0$ and $\alpha^{\mu}_{\nu}= \alpha(x) \bar{\sigma}^{\mu}_{\nu}$ in the following.
Note that this assumption can be realized, e.g., in quartic Horndeski theories.
Under this assumption, the action determining the background dynamics is given by
\begin{equation}
\label{tadpole_action}
	S_{\rm tadpole}=\int{{\rm d}^{4}x \sqrt{-g} \left[ \frac{M^{2}(t,r)}{2}R - \Lambda(t,r) - c(t,r)g^{rr} - \tilde{\beta}(t,r) K - \alpha(t,r) \bar{K}^{\mu}_{\nu}K^{\nu}_{\mu} \right] }\;,
\end{equation}
where we have defined $\tilde{\beta}= \beta - \alpha \bar{K}/3$.
It should be noted that the EFT parameters are functions of $(t,r)$ due to the spherical symmetry of the background.
Note also that quadratic HOST theories can be simply accommodated by introducing a term~$\zeta(t,r)n^\mu \partial_\mu g^{rr}$, but we omit this term in the present paper for simplicity.
The Einstein equation derived from the action~(\ref{tadpole_action}) is
\begin{equation}\label{Einstein_Eq}
	\hat{E}_{\mu \nu} M^{2} = \bar{T}_{\mu \nu}\;,
\end{equation}
where we have defined the differential operator~$\hat{E}_{\mu\nu}$ as follows:
\begin{align}
    \label{Modified_Einstein_Eq}
    \hat{E}_{\mu \nu} \equiv \bar{G}_{\mu\nu}&- \bar{\nabla}_\mu\bar{\nabla}_\nu+\bar{h}_{\mu\nu}\bar{\Box}
    -\bar{g}_{\mu\nu}\bar{n}^\lambda  \bar{n}^\sigma\bar{\nabla}_\lambda\bar{\nabla}_\sigma
    +2\bar{n}^\lambda \bar{n}_{(\mu}\bar{\nabla}_{\nu)}\bar{\nabla}_\lambda \nonumber \\
    &+2\brb{\bar{n}_{(\mu}\big(\bar{K}^\lambda_{\nu)}-\bar{K}\bar{h}^\lambda_{\nu)}\big)+\bar{a}_{(\mu}\bar{h}^\lambda_{\nu)}-\bar{h}_{\mu\nu}\bar{a}^\lambda} \bar{\nabla}_\lambda\;,
\end{align}
with $a_\mu \equiv n^\lambda\nabla_\lambda n_\mu$ and $\bar{G}_{\mu\nu}$ the background Einstein tensor.
When $M^2$ is a constant, the left-hand side of Eq.~\eqref{Einstein_Eq} is simply given by $M^2\bar{G}_{\mu\nu}$.
Note also that the operator~$\hat{E}_{\mu\nu}$ can also be written as
    \begin{align}
    \hat{E}_{\mu \nu} = \bar{G}_{\mu\nu}-\bar{\rm D}_\mu\bar{\rm D}_\nu+\bar{h}_{\mu\nu}\bar{\rm D}_\lambda\bar{\rm D}^\lambda
    +\brb{2\bar{n}_{(\mu}\big(\bar{K}^\lambda_{\nu)}-\bar{K}\bar{h}^\lambda_{\nu)}\big)
    -\big(\bar{K}_{\mu\nu}-\bar{K}\bar{h}_{\mu\nu}\big)\bar{n}^\lambda
    +2\big(\bar{a}_{(\mu}\bar{h}^\lambda_{\nu)}-\bar{h}_{\mu\nu}\bar{a}^\lambda\big)}\bar{\nabla}_\lambda\;,
\end{align}
where ${\rm D}_\mu$ stands for the covariant derivative on a timelike hypersurface of constant $\Phi$.
The background stress-energy tensor introduced in Eq.~(\ref{Einstein_Eq}) is given by 
\begin{align}
	\bar{T}_{\mu \nu}  = & -\bigl(c \bar{g}^{r r} + \Lambda + \alpha \bar{K}^{\lambda}_{\sigma} \bar{K}^{\sigma}_{\lambda}\bigr) \bar{g}_{\mu \nu} + 2 c \delta^{r}_{\mu} \delta^{r}_{\nu}  - 2 \bar{n}_{( \mu}\partial_{\nu)}\tilde{\beta} + \bigl(\bar{g}_{\mu \nu}+\bar{n}_{\mu} \bar{n}_{\nu}\bigr) \bar{n}^{\lambda} \partial_{\lambda} \tilde{\beta} \nonumber \\ & + 2 \alpha \bar{K}^{\lambda}_{\mu} \bar{K}_{\lambda \nu} - \alpha \bar{K}^{\lambda}_{\sigma} \bar{K}^{\sigma}_{\lambda} \bar{n}_{\mu} \bar{n}_{\nu} - 2 \bar{\nabla}_{\lambda}(\alpha \bar{K}^{\lambda}_{ ( \mu} \bar{n}_{\nu )} ) + \bar{\nabla}_{\lambda}(\alpha \bar{K}_{\mu \nu} \bar{n}^{\lambda})\; .
\end{align}
Notice that the expression for $\bar{T}_{\mu\nu}$ above differs from that in the case of a timelike scalar profile~\cite{Mukohyama:2022enj,Mukohyama:2022skk} due to the change in the nature of $n^\mu$, which is spacelike here rather than timelike.

Using the ansatz~(\ref{eq:bg_metric_Le}) for the metric, the Einstein equation reduces to
\begin{equation}\label{eq:ein_cont_M2}
    \begin{aligned}
    B c - \Lambda &= M^2 \bar{G}^r{}_{r}\;, \\
    -\Lambda - cB - \frac{B F'}{F}(M^2)' + \frac{A'B \alpha'}{2A}
    + \sqrt{B}\,\tilde{\beta}' + H_{1} \alpha + \frac{\dot{F}}{AF} (M^2)^{\boldsymbol{\cdot}} &= M^2 \bar{G}^t{}_{t}\; ,\\
    H_2 \alpha + \frac{A'\dot{\alpha} }{2A^2} + \frac{\dot{\tilde{\beta}}}{A \sqrt{B}}  - \frac{F'}{A F} (M^2)^{\boldsymbol{\cdot}} &= M^2 \bar{G}^t{}_{r} \;, \\
    H_3 (\alpha + M^2)' + H_{4} \alpha   - (M^2)^{\boldsymbol{\cdot}} \bigg[\frac{(F A)^{\boldsymbol{\cdot}}}{2 A^2 F} + \frac{\dot{B}}{A B}\bigg] + \frac{(M^2)^{\boldsymbol{\cdot}\boldsymbol{\cdot}}}{A}&= M^2 (\bar{G}^\theta{}_{\theta} - \bar{G}^t{}_{t}) \;,
    \end{aligned}
\end{equation}
where a dot denotes a derivative with respect to $t$, a prime a derivative with respect to $r$, and we have defined
\begin{equation}
 \begin{aligned}
     H_1 &\equiv -\frac{1}{4A} \bigg[2B\bigg(\frac{A'^2}{A} + \frac{A F'^2}{F^2}\bigg) - A'B' \bigg(1 + \frac{2BF'}{F B'}\bigg) - 2 BA''\bigg] \;, \\
    H_2 &\equiv \frac{1}{4} \bigg[\bigg(\frac{A'}{A^2}\bigg)^{\boldsymbol{\cdot}} + \frac{(BA')^{\boldsymbol{\cdot}}}{BA^2} + \frac{2 \dot{F}}{A F}\bigg(\frac{A'}{A} -  \frac{F'}{F} \bigg)\bigg] \;, \\
     H_3 &\equiv \frac{B}{2} \bigg(\frac{F'}{F} - \frac{A'}{A}\bigg) \;, \qquad H_4 \equiv \frac{B}{4}\bigg[\bigg(\frac{A'}{A} - \frac{F'}{F} \bigg)\bigg(\frac{A'}{A} - \frac{B'}{B}\bigg) - \frac{2 A''}{A} + \frac{2 F''}{F}\bigg] \;.
 \end{aligned}
 \end{equation}
The non-vanishing components of the Einstein tensor~$\bar{G}^\mu{}_{\nu}$ are given by 
\begin{equation}
     \begin{aligned}
         \bar{G}^t{}_{t} &= -\frac{1}{F}\bigg(1 - \frac{B' F'}{2} + \frac{B {F'}^2}{4 F} - B F''+\frac{\dot{B} \dot{F}}{2 A B} + \frac{\dot{F}^2}{4 A F}\bigg) \;, \\
         \bar{G}^r{}_{r} &= -\frac{1}{F}\bigg(1 - \frac{B A' F' }{2 A} - \frac{B {F'}^2}{4 F} - \frac{\dot{A} \dot{F}}{2 A^2} - \frac{\dot{F}^2}{ 4 A F} + \frac{\ddot{F}}{A}\bigg) \;, \\
         \bar{G}^t{}_{r} &= \frac{1}{2A F}\bigg(\frac{F'\dot{B}}{B}- \frac{A' \dot{F}}{A} - \frac{F'\dot{F}}{F} + 2 \dot{F'}\bigg) \;, \\
         \bar{G}^\theta{}_{\theta} &=    \bar{G}^\phi{}_{\phi} =  \frac{BA'}{4A}\bigg(\frac{B'}{B} - \frac{A'}{A} + \frac{F'}{F} \bigg)  + \frac{BF'}{4F}\bigg(\frac{B'}{B} - \frac{F'}{F} \bigg) + \frac{\dot{F}}{4AF}\bigg(\frac{\dot{A}}{A} + \frac{\dot{F}}{F}\bigg)      \\
         & \hspace{4.5mm} + \frac{\dot{B}}{4AB}\bigg(\frac{\dot{F}}{F} - \frac{\dot{A}}{A} - \frac{3 \dot{B}}{ B}  \bigg)   + \frac{B}{2}\bigg(\frac{A''}{A} + \frac{F''}{F} \bigg) + \frac{1}{2A}\bigg(\frac{\ddot{B}}{B} - \frac{\ddot{F}}{F} \bigg) \;. 
     \end{aligned}
\end{equation}
It is important to point out that the above background equations generally apply to the case in which the scalar field~$\Phi$ is non-minimally coupled to gravity, i.e., $M^2$ is not constant.  
In addition, it is straightforward to verify that Eq.~(\ref{eq:ein_cont_M2}) is consistent with those found in \cite{Franciolini:2018uyq} when the metric is taken to be time-independent.\footnote{Since we have derived Eq.~(\ref{eq:ein_cont_M2}) using the 4d Ricci scalar with the boundary term subtracted, additional terms arise from integration by parts when compared to the background equations derived from the standard 4d Ricci scalar.}
In the next section, we will specifically use these background equations to study a black hole solution with a linearly time-varying mass.

For completeness, let us discuss the compatibility between the consistency relation~(\ref{eq:EFT_consis_1}) and the background equations~(\ref{eq:ein_cont_M2}).
Using the background metric~(\ref{eq:bg_metric_Le}), the consistency relation~\eqref{eq:EFT_consis_1} yields
\begin{align}
\label{constraint_consistency_general_background}
    \dot{\Lambda}+\dot{c}\bar{g}^{rr}+\dot{\tilde{\beta}}\bar{K}+(\alpha\bar{K}^\mu_\nu)^{\boldsymbol{\cdot}}\bar{K}^\nu_\mu-\frac{1}{2}(M^2)^{\boldsymbol{\cdot}}\bar{R}=0 \;,
\end{align}
where we have disregarded the angular components of Eq.~(\ref{eq:EFT_consis_1}) since the EFT coefficients are now functions of $t$ and $r$.
Using the background equations~\eqref{eq:ein_cont_M2}, one can eliminate $\Lambda$, $c$, and $\tilde{\beta}$ from Eq.~\eqref{constraint_consistency_general_background}, leading to the following expression:
\begin{align}\label{eq:consis_con}
   \frac{\dot{F}}{F} \bigg[\mathcal{H}_1 M^2+ \mathcal{H}_2 (M^2)^{\boldsymbol{\cdot}} + \mathcal{H}_3 (M^2)' + \frac{1}{A} (M^2)^{\boldsymbol{\cdot}\boldsymbol{\cdot}}  + \mathcal{H}_4 \alpha + \mathcal{H}_5 \alpha'\bigg] = 0 \;, 
\end{align}
where we have defined 
\begin{equation}
    \begin{aligned}
    \mathcal{H}_1 &\equiv \frac{A'B}{4A}\bigg(\frac{A'}{A} -\frac{B'}{B} - \frac{F'}{F}\bigg) + \frac{\dot{B}}{4AB}\bigg(\frac{\dot{A}}{A}  + \frac{\dot{F}}{F} + \frac{3 \dot{B}}{B} \bigg) + \frac{1}{2A}\bigg( \frac{\ddot{F}}{F} - \frac{\ddot{B}}{B}\bigg)   \\ 
    &\hspace{4mm} - \frac{\dot{F}}{2AF}\bigg( \frac{\dot{A}}{2A} + \frac{\dot{F}}{F}\bigg) + \frac{B}{2}\bigg(\frac{F''}{F}-\frac{A''}{A }\bigg) + \frac{B' F'}{4 F} -\frac{1}{F} \;, \\
    \mathcal{H}_2 &\equiv  - \frac{1}{A}\bigg( \frac{\dot{A}}{2 A} + \frac{\dot{B}}{B} + \frac{\dot{F}}{2F} \bigg)\;, \qquad \mathcal{H}_3 \equiv \frac{B}{2}\bigg(\frac{F'}{F}-\frac{A' }{A}\bigg) \;, \\
   \mathcal{H}_4 &\equiv \frac{B}{4}\bigg(\frac{F'}{F} - \frac{A'}{A}\bigg)\bigg(\frac{B'}{B} - \frac{A'}{ A}\bigg)+ \frac{B}{2}\bigg(\frac{F''}{F} -\frac{A''}{A} \bigg) \;, \qquad
   \mathcal{H}_5 \equiv \frac{B}{2}\bigg(\frac{F'}{F}-\frac{A'}{A}\bigg) \;.
\end{aligned}
\end{equation}
We see that, in the case where $F(t,r) = r^2$, the consistency relation~(\ref{eq:consis_con}) is automatically satisfied, regardless of the metric functions~$A(t,r)$ and $B(t,r)$.
If, on the other hand, the function~$F(t,r)$ acquires time dependence, Eq.~(\ref{eq:consis_con}) provides a non-trivial relation between $M^2(t,r)$ and $\alpha(t,r)$.
Additionally, similarly to Eq.~(\ref{constraint_consistency_general_background}), it is straightforward to express Eqs.~(\ref{eq:EFT_consis_2}) and (\ref{eq:EFT_consis_3}) using the ansatz~(\ref{eq:bg_metric_Le}) and the background equations~(\ref{eq:ein_cont_M2}).
The resulting consistency relations, as expected, yield non-trivial relations between the tadpole parameters and those associated with second-order terms in the EFT action.

Let us now consider Eq.~(\ref{eq:ein_cont_M2}) and see how we can express $\Lambda(t,r)$, $c(t,r)$, $\tilde{\beta}(t,r)$, and $\alpha(t,r)$ in terms of the metric functions.
For simplicity, we assume $F(t,r) = r^2$ and $M^2(t,r) = M_\star^2 = const$.
Note that the latter assumption can be achieved via a conformal transformation. 
Also, in principle, this assumption can be generalized by taking into account small perturbations, i.e., $F(t,r) = r^2[1 + \epsilon f_1(t,r)]$ and $M^2(t,r) = M_\star^2[1 + \epsilon f_2(t,r)]$ where $|\epsilon| \ll 1$ and $f_1,f_2$ are bounded functions of at most order unity.
Under the assumptions $F(t,r) = r^2$ and $M^2(t,r) = M_\star^2 = const$, the first two equations of (\ref{eq:ein_cont_M2}) yield
\begin{align}\label{eq:Lambda_exp}
	\Lambda(t,r) &= \frac{M_{\star}^{2}}{r^2} \left(1 - B - \frac{r B A'}{2 A} - \frac{rB'}{2} \right) - \frac{B\alpha}{r^2} \left[1 
    - \frac{rA'}{2A}\bigg(1 - \frac{rA'}{2 A} + \frac{rB'}{4 B}  \bigg)  - \frac{r^2A''}{4 A}\right] \nonumber \\
    & \hspace{5mm} + \frac{B A' }{4 A}\alpha' + \frac{\sqrt{B}}{2}  \tilde{\beta}'\;,
\end{align}
and
\begin{align}\label{eq:c_exp}
	c(t,r) = \frac{M_{\star}^{2}}{2r} \left( \frac{A'}{A} - \frac{B'}{B} \right)  - \frac{\alpha}{r^2} \left[1 
    - \frac{rA'}{2A}\bigg(1 - \frac{rA'}{2A} + \frac{rB'}{4B} \bigg) - \frac{r^2A''}{4 A}\right] + \frac{A'\alpha'}{4 A} + \frac{\tilde{\beta}' }{2\sqrt{B}}\;,
\end{align}
respectively, where we have used the $(tr)$-component of the background equations~(\ref{eq:ein_cont_M2}) to replace $\dot{\alpha}$ with other parameters and their derivatives. 
From the above expressions, we see that both $\Lambda(t,r)$ and $c(t,r)$ are solely determined in terms of $\alpha$ and $\tilde{\beta}$ as well as the metric functions.
Below, we will explicitly write down the differential equations for $\alpha(t,r)$ and $\tilde{\beta}(t,r)$.

Using the same assumption, the third and the forth equations of (\ref{eq:ein_cont_M2}) give
\begin{align}\label{eq:beta_tilde}
	-\frac{M_{\star}^2 \dot{B}}{r B} 
	+ \frac{A' \dot{\alpha}}{2 A} + \frac{\dot{\tilde{\beta}}}{\sqrt{B}}
	- \frac{\alpha}{2A} \left[A'\bigg(\frac{ \dot{A}}{ A} 
	- \frac{\dot{B}}{2 B}\bigg) 
	-  \dot{A}' \right]  = 0 \;,
 \end{align}
and 
\begin{align}
	\left(1 - \frac{r A'}{2 A}\right) r B \alpha' + B(M_{\star}^2+ \alpha) 
    \left[1 + \frac{r}{2}\bigg(\frac{B'}{B} - \frac{A'}{A}\bigg)\bigg(1 - \frac{r A' }{2 A }\bigg)  - \frac{r^2A''}{2 A} \right]& \nonumber \\
	- M_{\star}^2 \left[1 - \frac{r^2\dot{B}}{4AB}\bigg(\frac{\dot{A}}{A } +  \frac{3 \dot{B}}{ B}\bigg) + \frac{r^2 \ddot{B}}{2 A B} \right] &= 0\; , \label{eq:alpha_A_neq_B}
\end{align}
respectively.
From the above equations, we see that once $\alpha(t,r)$ is determined using Eq.~(\ref{eq:alpha_A_neq_B}), we can use that solution in Eq.~(\ref{eq:beta_tilde}) to solve for $\tilde{\beta}(t,r)$.

As an illustrative example, let us consider a static background metric satisfying the condition~$A(r) = B(r)$.
In this case, Eq.~(\ref{eq:alpha_A_neq_B}) can be analytically solved for $\alpha(r)$ as follows:
\begin{align}\label{eq:alpha_static_sol}
    \alpha(r) = \frac{M_\star^2(2 -2 A + r A')}{2A - rA'} + \frac{\lambda}{r(2A - r A')} \;,
\end{align}
where $\lambda$ is a constant of integration. 
Also, Eq.~(\ref{eq:beta_tilde}) gives $\dot{\tilde{\beta}} = 0$, implying that $\tilde{\beta}$ is an arbitrary function of $r$.
By use of the identity~\eqref{beta_redef}, $\tilde{\beta}=\tilde{\beta}(r)$ can be absorbed into other EFT parameters via integration by parts, and therefore one can set $\tilde{\beta}=0$ without loss of generality.
Then, using both solutions for $\alpha(r)$ and $\tilde{\beta}(r)\,(=0)$ in Eqs.~(\ref{eq:Lambda_exp}) and (\ref{eq:c_exp}), we can straightforwardly obtain the expression for $\Lambda(r)$ and $c(r)$.
In addition, for the Schwarzschild metric where $A(r) = 1 - r_{\rm s}/r$ with $r_{\rm s}$ corresponding to the Schwarzschild radius, Eq.~(\ref{eq:alpha_static_sol}) yields\footnote{In the case of a spacelike scalar profile, it is expected that the strong-coupling problem of perturbations around stealth backgrounds cannot be avoided by including the so-called scordatura term~\cite{Motohashi:2019ymr}.}
\begin{align}\label{eq:alpha_static_Sch}
    \alpha(r) = \frac{\lambda + 3 r_{\rm s} M_\star^2}{2r - 3r_{\rm s}} \;.
\end{align}
This implies that $\alpha$ is singular at $r=3r_{\rm s}/2$ so long as $\lambda+3r_{\rm s}M_\star^2\ne 0$.
Once we remove the singularity by setting $\lambda=-3r_{\rm s}M_\star^2$, $\alpha(r)$ identically vanishes.
Note that the condition~$\alpha=0$ is automatically satisfied in cubic Horndeski theories (see the dictionary in \cite{Franciolini:2018uyq}).
We emphasize that the above conclusion is based on Eq.~(\ref{eq:alpha_static_Sch}), which holds specifically for the stealth Schwarzschild background.

\section{Black hole with linearly time-varying mass}\label{sec:example_slow_BH}
In this section, as an application of our EFT, we study a concrete example where the metric functions~$A(t,r)$ and $B(t,r)$ in Eq.~(\ref{eq:bg_metric_Le}) are given by
\begin{align}\label{eq:metric_slow_mass}
    A(t,r) = B(t,r) = 1 - \frac{r_c(t)}{r} \;, 
\end{align}
where the apparent horizon radius~$r_c(t)$ is for simplicity assumed to be a linear function of $t$ in the time interval of interest:
\begin{align}
    r_c(t) = r_{\rm s} +  2 \dot{m} (t-t_0) \;,
\end{align}
with $r_{\rm s}$, $\dot{m}$, and $t_0$ being constants. 
In this section, we assume $F(t,r) = r^2$ and $M^2(t,r) = M_\star^2 = const$ for simplicity. Then, using Eq.~(\ref{eq:metric_slow_mass}) in (\ref{eq:alpha_A_neq_B}), we have
\begin{align}\label{eq:diff_eq_alpha}
    \bigg(r- \frac{3}{2}r_c\bigg)\alpha' + \alpha + \frac{4 \dot{m}^2  M_\star^2 r^3}{(r-r_c)^3} = 0 \;. 
\end{align}
The equation above can be analytically solved as 
\begin{align}\label{eq:sol_alpha_exact}
    \alpha(t,r) = \frac{\tilde{F}(t-t_0)}{2r - 3r_c} - 4  \dot{m}^2 M^2_\star 
 \bigg\{\frac{6 {r_c} ( r - r_c) + r^2}{{(r-r_c)}^2} + \frac{6 r_c}{2 r - 3 r_c}\log\bigg[ 2 \bigg(\frac{r}{r_c} - 1\bigg)\bigg]\bigg\} \;,
\end{align}
where $\tilde{F}(t-t_0)$ is an arbitrary function of $t - t_0$. 
Notice that the solution for $\alpha(t,r)$ above has been obtained without expanding in the small parameter~$|2 \dot{m} (t-t_0)|/r_{\rm s}$. 
From Eq.~(\ref{eq:sol_alpha_exact}), the apparent singularity at $r = 3r_c/2$ can be removed by setting $\tilde{F}(t-t_0) = 0$.
In this case, we have
\begin{align}\label{eq:reg_alpha}
     \alpha(t,r) = - 4  \dot{m}^2 M^2_\star 
 \bigg\{\frac{6 {r_c} ( r - r_c) + r^2}{{(r-r_c)}^2} + \frac{6 r_c}{2 r - 3 r_c}\log\bigg[ 2 \bigg(\frac{r}{r_c} - 1\bigg)\bigg]\bigg\} \;,
\end{align}
for which we have $\alpha(t,r) \to -108 \dot{m}^2 M_\star^2$ in the limit of $r \to 3r_c/2$.
We note that, in the static limit where $\dot{m} = 0$, the regularized $\alpha(t,r)$ vanishes, which is consistent with the discussion associated to  Eq.~(\ref{eq:alpha_static_Sch}). 
In addition, in the asymptotic region~$r/r_c \to \infty$, we have $\alpha(t,r) \simeq - 4  \dot{m}^2 M^2_\star + \mathcal{O}(r^{-1})$.
In the following, we will use the regularized solution for $\alpha(t,r)$ given by Eq.~(\ref{eq:reg_alpha}) to solve for the remaining tadpole parameters.

Let us now consider Eq.~(\ref{eq:beta_tilde}).
In this setup this equation reduced to
\begin{align}
   \sqrt{1- \frac{r_c}{r}}\, \dot{\tilde{\beta}}  + \frac{r_c\dot{\alpha}}{2r^2} + \frac{\, \dot{m}(2r - r_c)\alpha}{2r^2(r-r_c)} + \frac{2 \,\dot{m}M_\star^2}{r^2} = 0\;.
\end{align}
Given the solution for $\alpha(t,r)$ in Eq.~(\ref{eq:reg_alpha}), the above equation can be analytically solved as
\begin{align}\label{eq:sol_beta_exact0}
    \tilde{\beta}(t,r)
    &=\tilde{G}(r)+\frac{2M_\star^2}{r}\left(\sqrt{1-\frac{r_c}{r}}-\sqrt{1-\frac{r_{\rm s}}{r}}\right)-\frac{r_c\alpha}{2r^{3/2}\sqrt{r-r_c}}\;,
\end{align}
where $\tilde{G}(r)$ is an arbitrary function of $r$.
Notice that, similarly to the solution~\eqref{eq:sol_alpha_exact} for $\alpha$, this solution for $\tilde{\beta}(t,r)$ has been obtained without relying on a small parameter expansion.
Using Eq.~\eqref{beta_redef}, $\tilde{G}(r)$ can be absorbed into $\Lambda$ and $c$, and therefore its contribution can be removed from $\tilde{\beta}$. Hence, without loss of generality, we have
\begin{align}\label{eq:sol_beta_exact}
    \tilde{\beta}(t,r)
    &=\frac{2M_\star^2}{r}\left(\sqrt{1-\frac{r_c}{r}}-\sqrt{1-\frac{r_{\rm s}}{r}}\right)-\frac{r_c\alpha}{2r^{3/2}\sqrt{r-r_c}}\;.
\end{align}
We see that $\tilde{\beta}(t,r)$ is regular for $r > \max(r_c,r_{\rm s})$, provided $\alpha(t,r)$ is regularized as in Eq.~(\ref{eq:reg_alpha}).
It should also be noted that $\tilde{\beta}$ vanishes in the static limit where $\dot{m}=0$, which is consistent with the discussion below Eq.~\eqref{eq:alpha_static_sol}.\footnote{In Eq.~\eqref{eq:sol_beta_exact0}, the $r$-dependent integration constant~$\tilde{G}(r)$ has been chosen such that $\tilde{\beta} \to \tilde{G}$ in the static limit~$\dot{m} \to 0$, but an ambiguity still remains up to terms that vanish in the static limit.}
Given the solutions for $\alpha(t,r)$ and $\tilde{\beta}(t,r)$ [Eqs.~(\ref{eq:reg_alpha}) and (\ref{eq:sol_beta_exact}), respectively], the expressions for $\Lambda(t,r)$ and $c(t,r)$ from Eqs.~(\ref{eq:Lambda_exp}) and (\ref{eq:c_exp}) are
\begin{align}
    \Lambda(t,r) &= -\frac{\alpha}{r(r - r_c)}\bigg(1 - \frac{2r_c}{r} + \frac{9r_c^2}{8r^2}\bigg) + \frac{r_c \alpha'}{4r^2} + \frac{\tilde{\beta}'}{2}\sqrt{1-\frac{r_c}{r}} \;, \\
    c(t,r) &= -\frac{\alpha}{(r-r_c)^2}\bigg(1 - \frac{2r_c}{r} + \frac{9r_c^2}{8r^2}\bigg) + \frac{r_c \alpha'}{4r(r-r_c)} + \frac{r\tilde{\beta}'}{2(r-r_c)}\sqrt{1-\frac{r_c}{r}} \;.
\end{align}
It should be noted that there is a relation~$\Lambda=(1-r_c/r)c$.
The expressions above are also regular in the range~$r > \max(r_c,r_{\rm s})$ and are valid to all orders in $|2 \dot{m} (t-t_0)|/r_{\rm s}$.
Finally, these solutions of the EFT parameters will be particularly useful for analyzing the dynamics of perturbations and identifying the regime in which the theory becomes strongly coupled.
We leave these studies to future work.

Before closing this section, we emphasize that the EFT formulated in this paper applies to generic background geometries with the assumption that $\partial_\mu \bar{\Phi}$ is non-vanishing and spacelike everywhere in a spacetime region of interest. 
As discussed before, this implies the EFT can break down in the asymptotic region where the background of $\Phi$ becomes trivial. 
At this stage, there are two possible options one can proceed.
The first option is to consider a separate EFT around the trivial background in the asymptotic regime and match it properly with our EFT around black hole backgrounds.
In this case, the matching conditions serve as boundary conditions for solving perturbations within our EFT framework, thereby affecting observables such as the quasinormal-mode spectrum. 
Another approach is to argue that both the perturbative and derivative expansions are valid (and hence the EFT is still applicable) in the asymptotic region, which amounts to showing that the cutoff scale approaches zero more slowly than the derivatives and the size of perturbations. 
This, of course, requires a more detailed and delicate analysis, which we leave for future study.

\section{Conclusions}\label{sec:conc}
We have formulated the effective field theory (EFT) of scalar-tensor gravity on arbitrary backgrounds with a \textit{spacelike} scalar profile. 
The formulation of the EFT relies on the fact that in the unitary gauge a radial-dependent background of the scalar field, $\bar{\Phi}(r)$, spontaneously breaks the radial diffeomorphism invariance, while the EFT remains invariant under the time and angular diffeomorphisms, referred to as the $(2+1)$d diffeomorphism symmetry. 
It should be noted that the EFT of black hole perturbations with a spacelike scalar profile on a static and spherically symmetric background was first developed in \cite{Franciolini:2018uyq}, and was later extended in \cite{Hui:2021cpm} to accommodate a slowly rotating black hole background.  
In this paper, we have generalized such an EFT framework to arbitrary black hole backgrounds. 
In Section~\ref{sec:uni_gen_EFT}, we have provided a generic EFT action in the unitary gauge, which is a spacetime integral of a scalar function built from $(2+1)$d diffeomorphism covariant quantities, for instance, the $(2+1)$d curvature, the extrinsic curvature, and $g^{rr}$. 
This construction is similar to that of the EFT of perturbations with a timelike scalar profile in \cite{Mukohyama:2022enj,Mukohyama:2022skk}.
In particular, we have pointed out that, when expanded around non-trivial backgrounds, each individual term of the EFT action explicitly breaks the $(2+1)$d diffeomorphism invariance.
This is because the background values of the EFT building blocks (e.g., $\bar{g}^{rr}$ and $\bar{K}$) generally depend on both space and time. 
In order for the EFT to be invariant under the $(2+1)$d diffeomorphisms, it is necessary to impose a set of consistency relations among the expansion coefficients which ensures the $(2+1)$d diffeomorphism invariance of the EFT at any order in perturbations.
In Section~\ref{sec:Consistency}, we have explicitly derived such consistency relations by applying the chain rule associated with the radial derivative to the Taylor-expansion coefficients, and in Section~\ref{sec:shift-sym_EFT}, we have obtained another set of consistency relations by requiring that the EFT remains invariant under the shift of the scalar field, $\Phi \to \Phi + const$.\footnote{Indeed, this is similar to that found in the shift-symmetric EFT of inflation~\cite{Finelli:2018upr}, and in the EFT of black hole perturbations~\cite{Khoury:2022zor}.} 
Moreover, in Section~\ref{sec:EFT_consis_action}, we have obtained the EFT action of perturbations in the unitary gauge with its coefficients satisfying the consistency relations.
In Section~\ref{sec:bg_dynamics}, we have derived the background equations of motion (or equivalently, the tadpole cancellation conditions) from the action~(\ref{tadpole_action}) on a spherically symmetric background metric~\eqref{eq:bg_metric_Le} that is not necessarily static.
These background equations will be particularly useful when analyzing the dynamics of perturbations around the background~(\ref{eq:bg_metric_Le}).
In Section~\ref{sec:example_slow_BH}, we have presented a concrete example by studying the background dynamics of a Schwarzschild black hole with a linearly time-varying mass and obtained analytic expressions for the EFT parameters [i.e., $\Lambda(t,r)$, $c(t,r)$, $\tilde{\beta}(t,r)$, and $\alpha(t,r)$].

There are several directions for future investigation.
First, as briefly discussed in the previous sections, the EFT we have constructed can break down in the asymptotic region where the background scalar field approaches a trivial configuration, i.e., $\bar{\Phi}'(r) \to 0$.
In this case, it becomes necessary to investigate how to consistently match the EFT formulated around black hole backgrounds to that around a trivial background.
Also, this would shed light on how the boundary conditions are properly imposed in the asymptotic region, with potential implications for observables such as the spectrum of quasinormal modes and the tidal Love numbers. 
Second, given that our EFT is applicable to generic backgrounds, an interesting avenue would be to apply it to stationary and axisymmetric black holes and to study the dynamics of perturbations around such spacetimes.
Finally, it is also worthwhile to use our EFT framework to explore higher-dimensional models---such as braneworld scenarios~\cite{Goldberger:1999uk,Bazeia:2008zx,Dzhunushaliev:2009va}---where the background of a bulk scalar field is spacelike.
We leave these issues for future studies.

\section*{Acknowledgements}
This work was supported in part by World Premier International Research Center Initiative (WPI), MEXT, Japan.
S.M.~was supported in part by JSPS (Japan Society for the Promotion of Science) KAKENHI Grant No.\ JP24K07017.
K.T.~was supported in part by JSPS KAKENHI Grant No.\ JP23K13101.
The work of V.Y.~is supported by grants for development of new faculty staff, Ratchadaphiseksomphot Fund, Chulalongkorn University and by the NSRF via the Program Management Unit for Human Resources \& Institutional Development, Research and Innovation Grant No.\ B39G680009.


\appendix

\section{Background quantities}\label{AppA:background}
\subsection{Schwarzschild solution}\label{AppA:background_Sch}
Here, as an example, we express the background quantities such as $\bar{K}^\mu_\nu$ and ${}^{(3)}\!\bar{R}^\mu_\nu$ for the Schwarzschild solution:
\begin{align}
    \bar{g}_{\mu\nu}{\rm d}x^\mu{\rm d}x^\nu = -A(r) {\rm d}t^2 + \frac{{\rm d}r^2}{A(r)} + r^2 \left({\rm d}\theta^2 + \sin^2\theta\, {\rm d}\phi^2\right) \;,
\end{align}
where the function~$A(r)$ is given by
\begin{align}
    A(r) = 1 - \frac{r_{\rm s}}{r} \;,
\end{align}
and $r_{\rm s}$ is the Schwarzschild radius.
Clearly, this solution is invariant under translations in time and angular coordinates. 
In this case, we have 
\begin{align}
    \bar{g}^{rr} = A \;, \qquad \bar{K} = \frac{1 + 3A}{2r \sqrt{A}} \;, \qquad \bar{\sigma}^t_t = -2\bar{\sigma}^\theta_\theta  = -2 \bar{\sigma}^\phi_\phi = \frac{1-3A}{3r\sqrt{A}} \;,
\end{align}
and 
\begin{align}
    {}^{(3)}\!\bar{R} = -3 \bar{\tilde{r}}^t_t  = 6 \bar{\tilde{r}}^\theta_\theta = 6 \bar{\tilde{r}}^\phi_\phi = \frac{2}{r^2}  \;.
\end{align}
We see that even in this simple example the background values of $\sigma^\mu_\nu$ and $\tilde{r}^\mu_\nu$ are non-vanishing.  

\subsection{General metric ansatz}\label{AppA:background_general}
Finally, we give the expressions for the background quantities of the ansatz~(\ref{eq:bg_metric_Le}).
For the trace of the extrinsic curvature and the non-vanishing components of $\bar{\sigma}^\mu_\nu$, we have  
\begin{align}
    \bar{K} &= \frac{\sqrt{B}}{2}\bigg(\frac{A'}{A} + \frac{2F'}{F}\bigg) \;, \qquad 
    \bar{\sigma}^t_t = -2\bar{\sigma}^\theta_\theta  = -2 \bar{\sigma}^\phi_\phi  = \frac{\sqrt{B}}{3}\bigg(\frac{A'}{A} - \frac{F'}{F}\bigg) \;.
\end{align}
In addition, the $(2+1)$d Ricci scalar~${}^{(3)}\!\bar{R}$ and the non-vanishing components of $\bar{\tilde{r}}^\mu_\nu$ are 
\begin{align}
    \begin{split}
    {}^{(3)}\!\bar{R} &= \frac{2}{F} - \frac{\dot{A}\dot{F}}{A^2 F} - \frac{\dot{F}^2}{2 A F^2} + \frac{2\ddot{F}}{AF} \;, \\
    \bar{\tilde{r}}^t_t &= -2\bar{\tilde{r}}^\theta_\theta = -2\bar{\tilde{r}}^\phi_\phi = - \frac{1}{3}\bigg(\frac{2}{F} + \frac{\dot{A}\dot{F}}{2A^2 F} + \frac{\dot{F}^2}{AF^2} - \frac{\ddot{F}}{AF}\bigg) \;.
    \end{split}
\end{align}
We see that, for this generic solution, all background quantities are functions of both $t$ and $r$.
Therefore, any scalar functions made out of the perturbations such as $\delta g^{rr}$ and $\delta K$ are no longer invariant under the temporal diffeomorphisms.

\bibliographystyle{utphys}
\bibliography{bib_v4}

\end{document}